\documentclass{article}
\usepackage[T1]{fontenc} 
\usepackage[utf8]{inputenc} 
\usepackage{ismir,amsmath,cite,url}
\usepackage{graphicx}

\usepackage{color}

\usepackage{amssymb}
\usepackage{amssymb}
\usepackage[inline]{enumitem}
\usepackage{lineno}
\usepackage{caption}
\usepackage{subcaption}
\usepackage[export]{adjustbox}
\usepackage{float}

\def\lbd{}	

\title{On the Typicality of Musical Sequences}

\threeauthors
  {Mathias Rose Bjare} {Computational Perception, Johannes Kepler University Linz \\ {\tt mathias.bjare@jku.at}}
  {
  } {
  }
  {Stefan Lattner} {Sony Computer Science Laboratories\\ {\tt stefan.lattner@sony.com}}

\def\authorname{Mathias Rose Bjare and Stefan Lattner}

\usepackage[bookmarks=false,pdfauthor={\authorname},pdfsubject={\papersubject},hidelinks]{hyperref}
\usepackage{cleveref}

\begin{document}
\maketitle
\begin{abstract}
It has been shown in a recent publication that words in human-produced English language tend to have an information content close to the conditional entropy. In this paper, we show that the same is true for events in human-produced monophonic musical sequences. We also show how ``typical sampling'' influences the distribution of information around the entropy for single events and sequences.
\end{abstract}
%
\section{Introduction}\label{sec:introduction}
It has recently been shown in \cite{typicaldecoding} that human-produced natural English sentences are composed of words which encode information close to the expected information content (i.e., the entropy). 
Derived from this finding, the authors of \cite{typicaldecoding} also proposed a sampling strategy called ``typical sampling'' that produced sentences that, compared to other sampling methods, were deemed more natural by humans.
In this paper, we evaluate if the trend towards the expected information content can also be observed in music. To that end, we reproduce some of the analyses performed in \cite{typicaldecoding} and \cite{prob_qual_paradox} for musical events and sequences. More precisely, we show how the information of human-produced musical material is distributed around the model entropy. We also evaluate how typical sampling impacts these distributions and how it compares to conventional, ancestral sampling.

\section{Methods}
In the following, we recap Information Theory concepts coined in the seminal paper \cite{shannon}, 
which have been applied to music in \cite{meyer} to explain expectation and surprise. Let $p\left(x_{t}\vert x_{<t}\right)$ be the conditional probability of a symbol $x_{t}$ given the past $x_{<t}$ observed symbols and $q$ a model fitted to $p$, e.g., a neural network fitted with likelihood optimization. A symbol could represent a word in natural language or a musical event in music.
The conditional \textit{information content} (IC) is given by $IC\left(x_{t}\vert x_{<t}\right) = -\log{q\left(x_{t}\vert x_{<t}\right)}$ 
and is a measure of how surprising (i.e., high IC) or expected (i.e., low IC) the symbol $x_{t}$ is according to the model $q$, given the past $x_{<t}$ observations. Furthermore, the IC of a whole sequence is defined as $IC\left(x\right) = \sum_{t=0}^{\lvert x \rvert -1}  IC\left(x_{t}\vert x_{<t}\right)$ and the information density $ID(x) = IC(x) / \lvert x \rvert$.
We are also interested in the expected information or \textit{entropy} $H$ for the conditional symbol distribution and the sequence distribution.
\subsection{Typicality}
In \cite{typicaldecoding}, the authors find that in natural language sentences, the conditional word information is close to the expected conditional information, and in \cite{prob_qual_paradox} that the sequence information content is close to the (unconditional) entropy or that
\begin{align}
    \lvert \epsilon_{sym} \rvert &= \lvert H\left(x_{t}\vert x_{<t}\right) - IC\left(x_{t}\vert x_{<t}\right)\rvert  \label{eq:typ_evt},\text{ and }\\ 
    \lvert \epsilon_{seq} \rvert &= \lvert H\left(x\right) - IC\left(x\right)\rvert,  \label{eq:typ_seq}
\end{align}
respectively, are small. We are interested in testing if the results transfer from natural language sequences of words to human compositions represented as sequences of musical events. As the lengths of musical sequences can vary a lot, we hypothesize that $IC\left(x\right)$ (in \cref{eq:typ_seq}) has less relevance to music than the (length-normalized) \emph{information density}, where the $ID(x)$ of individual songs should be close to the expected $ID(x)$, resulting in a small
\begin{equation}
    \lvert \epsilon_{ID} \rvert = \vert \mathbb{E}\left[ID\left(x\right)\right] - ID\left(x\right) \rvert.
    \label{eq:typ_id}
\end{equation}

\subsection{Typical Sampling}
\label{seq:typical_sampling}
In \cite{typicaldecoding}, a sampling strategy is proposed where the least typical symbols\footnote{Symbols with highest value of $\epsilon_{sym}$ in \cref{eq:typ_evt}} of $q$ get pruned. The obtained samples are reported to be more human-like than other sampling strategies. In typical sampling, we initially identify the smallest set of most typical symbols $V$ s.t. $q(V|x_{<t}) \geq \tau$, 
\begin{figure*}[!ht]
    \centering
    \begin{subfigure}[t]{.495\textwidth}
        \centering
        \includegraphics[trim=10 5 40 19,clip,scale=.6]{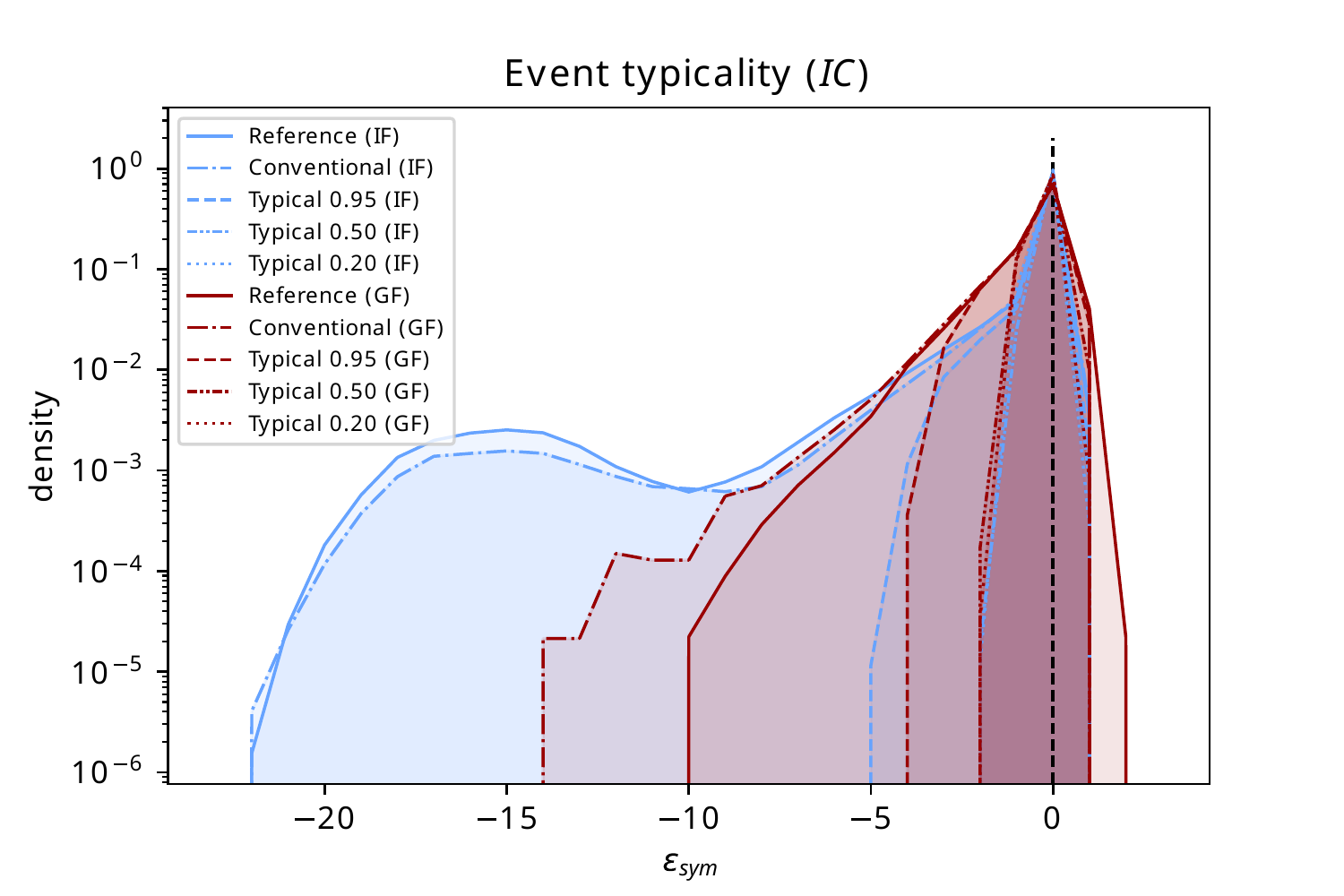}
        \caption{}
        \label{fig:evt_typ}
    \end{subfigure}
    \hfill
    \begin{subfigure}[t]{.495\textwidth}
        \centering
        \includegraphics[trim=15 5 35 19,clip,scale=.6]{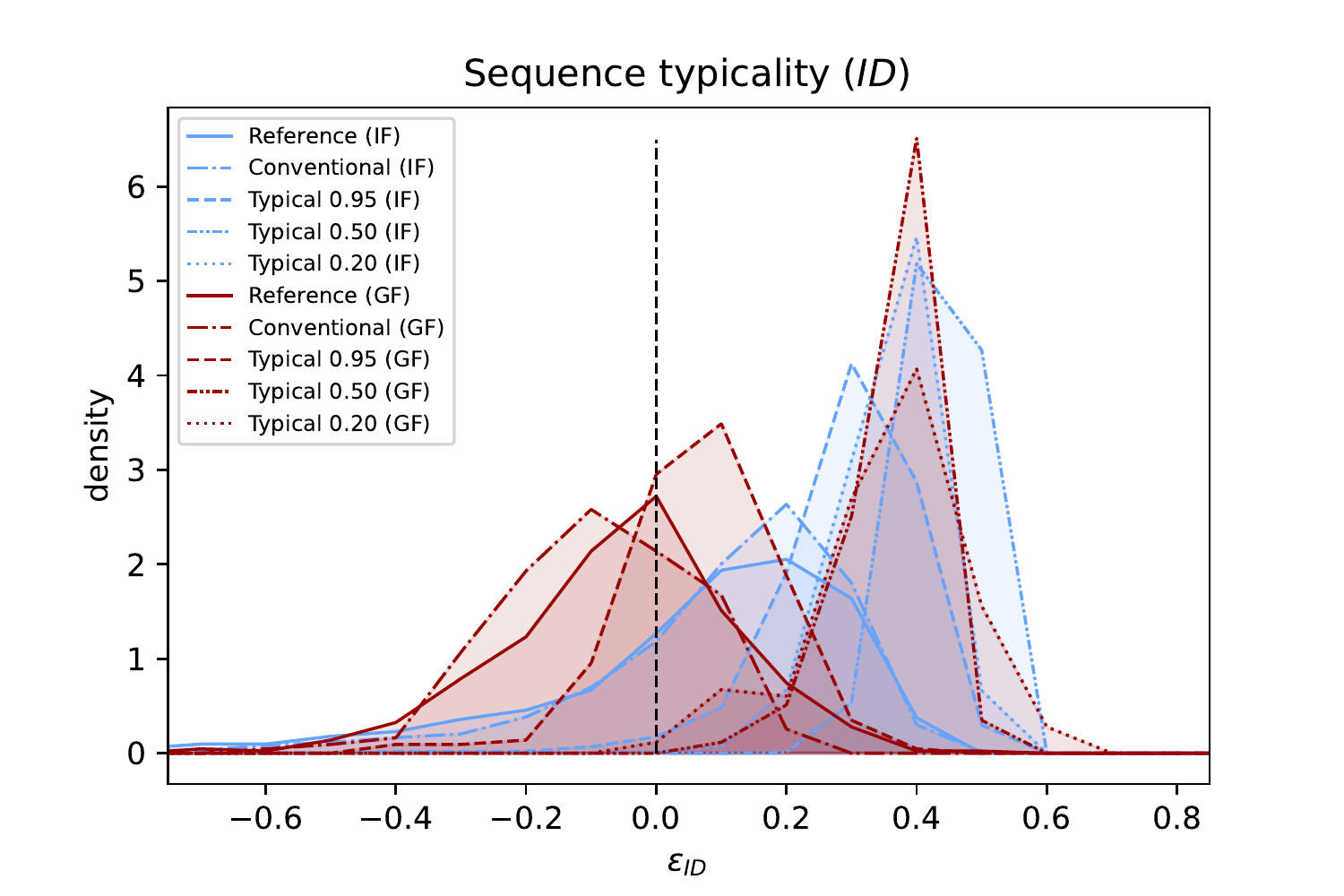}
        \caption{}
        \label{fig:seq_typ}
    \end{subfigure}
    \vspace{-3mm}
    \caption{Distribution of typicality divergence for human-composed music (Reference), conventional ancestral sampling (Conventional), and typical sampling (Typical) for single events (a) and sequences (b). The distributions are reported for datasets IF and GF.}
    \vspace{-1mm}
    \label{fig:results}
\end{figure*}

where $\tau$ determines the amount of probability pruned. Secondly, the probabilities of symbols $\notin V$ are zeroed. Finally, the resulting function is normalized to yield a valid probability distribution. We wish to investigate the effect of typical sampling on the typicality of generated sequences (\cref{eq:typ_evt,eq:typ_id}).

\section{Experiments}
Our experiments are performed on monophonic symbolic music. 
We use the dataset of \cite{folkrnnsession}, consisting of $45849$ Irish folk music lead sheets (denoted IF).
For computational reasons, we discard the $5\%$ longest sequences, resulting in a test set of $3618$ samples of length $\leq 691$.
In addition, we use all German folk songs from the Essen Folk Database \cite{essen1, essen2} (referred to as GF), consisting of 5152 melodies. 
We partition both datasets in training, validation, and test sets with proportions 10/12, 1/12, and 1/12, respectively (all analyses are performed on the test set).

The compositions are encoded using a simplified variant of the tokenization strategy described in \cite{performancernn}. We order the notes by absolute time and serialize a note as a note-value (pitch) or rest token, followed by one or more duration tokens indicating the note's duration. Specifically, we use $128$ \emph{change-pitch tokens} for pitches $C_{-1} - G_9$,
    $1$ \emph{rest token},
    $100$ \emph{duration tokens} quantized linearly from $10$ms to $1000$ms, and
    $1$ \emph{end-of-sequence token}.
This results in a vocabulary of 230 tokens. For each dataset, we train a Transformer decoder model \cite{attention} with relative attention \cite{relativeattention, musictransformer} in a self-supervised prediction task.
We sample sets of sequences from each model with a maximum sequence length and size following their respective test set's maximum length and size. We perform conventional ancestral sampling from the unmodified distribution and typical sampling with a low, medium, and high $\tau$ value. 

\section{Results and Discussion}
In \cref{fig:evt_typ}, we show the distributions of the information contents for single events relative to the entropy (for typical sampling, the entropy and information content of the events is always calculated based on the unpruned distributions). In \cref{fig:seq_typ}, we show the distributions of Information Densities (IDs) for sequences relative to the mean IDs (see \cref{eq:typ_id}) of the respective reference distribution. For both data sets, we show the distributions of the human-generated melodies (Reference), conventional ancestral sampling (Conventional), and typical sampling (Typical, with different $\tau$ values). 

Similar to \cite{typicaldecoding}, we find that the ICs of events are distributed densely around the conditional entropy (see \cref{fig:evt_typ}). When using typical sampling, this trend gets more pronounced for decreasing $\tau$ values, showing that the sampled musical events are indeed more typical.

However, also with conventional ancestral sampling, the highest density of the event's information is close to the entropy. In addition, the distributions of Conventional follow those of Reference closer than Typical in both figures. In \cref{fig:seq_typ}, typical sampling causes the ID distributions of generated sequences to narrow and shift to the right with decreasing $\tau$ values. Note that in \cref{fig:seq_typ}, $\epsilon$ is always calculated based on the entropy of the reference distributions, meaning that \emph{the generated sequences become more probable when $\tau$ decreases}. This is reasonable, as typical sampling prunes less likely events, as shown in \cref{fig:evt_typ}, where events to the left get pruned.
The Reference distribution of IF is not centred around $0$ in \cref{fig:seq_typ} (i.e., it is shifted to the right). This suggests that a considerable number of songs in this dataset have a high average ID (meaning that they are rather unlikely given the dataset distribution, causing the overall entropy to increase).

It is subject to future work (not shown in \cite{typicaldecoding}) to investigate if the typicality property of events is mainly linked to human-generated sequences. That is not certain, considering that even events sampled from a uniform or degenerate (i.e., deterministic) distribution are perfectly typical (with $\epsilon=0$). If it turns out that the typicality property is trivial, the reason why typical sampling yields sequences preferred by humans (as shown in \cite{typicaldecoding}) may also be explained by other factors, like a more stable random walk (as the model is confronted with more probable context), or by the simple fact that slightly higher probability sequences are more appealing.

\section{Conclusion}
We showed that musical events tend to be typical and that typical sampling generally yields more likely sequences than conventional sampling.
However, it is unclear if the typicality is a property of human-generated data only and if typicality of such sequences is a valid explanation for why typical sampling results in sequences preferred by humans.\let\thefootnote\relax\footnote{This work was conducted in a collaboration between JKU and  Sony Computer Science Laboratories Paris under a research agreement.}


\newpage

\bibliography{ISMIR2022_lbd}

%
%
%
%
%
\end{document}